\documentclass[aps,pra,twocolumn,amssymb,amsfonts,floatfix,showpacs]{revtex4}

\usepackage{graphicx,psfrag}
\usepackage{dcolumn}
\usepackage{bm}
\usepackage{mathbbol}
\begin{document}

\title{\mbox{Enhanced Convergence and Robust Performance of 
Randomized Dynamical Decoupling}}

\author{Lea F. Santos}
\email{Lea.F.Dos.Santos@Dartmouth.edu}
\author{Lorenza Viola}
\email{Lorenza.Viola@Dartmouth.edu}
\affiliation{\mbox{Department of Physics and Astronomy, 
Dartmouth College, 6127 Wilder Laboratory, Hanover, NH 03755, USA}}

\date{\today}

\begin{abstract}
We demonstrate the advantages of randomization in coherent quantum
dynamical control.  For systems which are either time-varying or
require decoupling cycles involving a large number of operations, we
find that simple randomized protocols offer superior convergence and
stability as compared to deterministic counterparts.  In addition, we
show how randomization may allow to outperform purely
deterministic schemes at long times, including combinatorial and
concatenated methods.  General criteria for optimally interpolating
between deterministic and stochastic design are proposed and
illustrated in explicit decoupling scenarios relevant to quantum
information storage.
\end{abstract}

\maketitle

Counterintuitive roles of randomness in physical processes have long
been acknowledged.  A paradigmatic example is stochastic
resonance~\cite{GammaSR}, whereby the transmission of a weak signal is
maximized at an optimum noise level.  Within quantum information
science, a number of applications benefiting from randomness in
quantum states and operations have been identified recently.
Suggestive evidence is offered by dissipation-assisted quantum
computation schemes~\cite{Beige} and by the idea that chaos may
stabilize quantum algorithms~\cite{Prosen}.  In the quantum
communication setting, the use of random operations has been shown to
decrease the communication cost of achieving remote state preparation
and of constructing efficient quantum data-hiding
schemes~\cite{Hayden}.  Finally, random unitary operators may pave the
way to efficient estimation methods for open quantum
systems~\cite{EmersonSc}.

In the context of quantum control theory, the fragility of quantum
coherence against {\em uncontrolled} noise and randomness hindered the
exploration of stochastic control methods till
recently~\cite{Mancini,Viola2005Random,Kern,Santos2005}.  General
control-theoretic arguments supporting the usefulness of {\em
controlled} randomness in dynamical decoupling (DD) schemes for
switching off unwanted evolution in interacting quantum systems were
provided in~\cite{Viola2005Random}, and have been validated in
specific examples~\cite{Kern} (see also~\cite{Santos2005}).  These
studies leave, however, several key questions largely unanswered.  In
particular: For given control resources, under what circumstances (if
any) can randomized DD match or outperform the {\em best}
deterministic schemes known to date?  Can randomized design be
exploited in general to further push the efficiency and viability of
DD schemes?
 
In this Letter, we present conclusive evidence of the benefits of
randomization in light of the above questions.  To isolate
the factors responsible for different DD performance, we focus on the
simplest control benchmark: complete refocusing of Hamiltonian
evolution
by means of a restricted but otherwise perfect set of
control operations.  
In the process, we verify conjectures made
in~\cite{Viola2005Random}, stretch the analysis into unexplored
domains, and offer criteria for constructing new highly efficient DD
protocols.  For time-independent systems, we demonstrate the {\em
superior averaging} of randomized protocols in control scenarios
involving sufficiently long control cycles and/or evolution times.  For
time-varying systems, randomized DD provides a {\em robust strategy}
in the presence of system uncertainties.  The combination of three key
ingredients -- concatenated control, symmetry, and randomization --
emerges as a general principle for design optimization.


{\em DD setting}.-- DD methods modify a target dynamics $H_0(t)$
through the addition of a control field $H_c(t)$.  We assume that
$||H_0(t)||_2=\max |{\rm eig} (H_0(t))| < \kappa$, for $\kappa >0$ and
all $t$.  Let $U_c(t) = {\cal T} \exp [- i\int_{0}^t H_c (u) du]$
($\hbar =1$) denote the control propagator, with ${\cal T}$ indicating
time ordering.  In a {\em logical} frame which explicitly removes
$H_c(t)$, the controlled evolution is described by the propagator
$\tilde{U}(t)=U_c^{\dagger}(t)U(t)={\cal T} \exp [- i\int _{0}^t
{\tilde H}(u) du]$, where $U(t)$ is the total propagator in the
physical frame and ${\tilde H}(t)=U^{\dagger}_c(t)H_0(t) U_c(t)$ is
the logical Hamiltonian~\cite{HaeberlenBook}.  If $H_0(t)$ is
time-independent and $U_c(t)$ periodic with a cycle time $T_c$, then
$\tilde{U}(T_n)= {U}(T_n)$, $T_n=nT_c$, $n\in {\mathbb N}$, and both
frames coincide stroboscopically. The DD objective here is to make
$\tilde{U}(T)$, for final time $T>0$, as close as possible to
$\openone$ according to a metric of choice.

Let the control resources be specified by a discrete set of unitary
operators corresponding to a (projective) representation of a group
${\cal G} = \{ {g}_j \}$, $j=0,\ldots,|{\cal G}|-1$, in the Hilbert
space ${\cal H}$ of the system, dim$({\cal H})=d < \infty$.  Both the
basic deterministic and random DD protocols may be understood as
effecting an appropriate symmetrization of $H_0(t)$ according to
${\cal G}$.  Assume a time-independent system first.  In standard {\em
deterministic DD}, average Hamiltonian theory (AHT) allows the logical
evolution to be represented as ${\tilde U}(T_n)=[\exp (-i \bar{H}
T_c)]^n$, where $\bar{H}=$$\sum_{k=0}^{\infty}\bar{H}^{(k)}$ and each
term $\bar{H}^{(k)}$ is computed from the Magnus
expansion~\cite{HaeberlenBook}.  Under the convergence condition
$\kappa T_c<1$, the leading zero-th order contribution $\bar{H}^{(0)}$
may be mapped into a group average $\bar{H}_{\cal G}={|{\cal
G}|^{-1}}\sum_{j} {g_j^\dagger H_0 g_j}$ by sequentially steering
$U_c(t)$ through {\em all} the group elements $\{ {g}_j \}$ -- which
corresponds to a train of ``bang-bang" instantaneous pulses $P_k=g_k
g_{k-1}^{\dagger}$, $k=1,\ldots, |{\cal G}|$, separated by $\Delta
t>0$ and with $T_c=|{\cal G}| \Delta t$~\cite{Viola1}.  A cyclic
protocol based on a {\em fixed}, predetermined control path within
${\cal G}$ will be referred to as periodic DD, henceforth ({\tt PDD}),
provided first-order averaging is achieved, ${\bar H}^{(0)}=0$.  The
simplest randomized version of such a DD procedure is obtained by
sampling $U_c(t)$ over ${\cal G}$ uniformly at random according to the
Haar measure~\cite{Viola2005Random} -- leading to what we call {\em
na\"{\i}ve random} DD ({\tt NRD}).  Let ${\mathbb E} \{\cdot\}$ denote
expectation over all possible control realizations; ${\cal
G}$-symmetrization still emerges {\em on average} at each $t$ via a
quantum operation ${\mathbb E}\{ U^\dagger_c(t) H_0 (t) U_c(t)\}=
\bar{H}_{\cal G}$, and convergence may be rigorously established in
the limit $T \Delta t \kappa^2 \ll 1$~\cite{Viola2005Random}.

Lower bounds for the expected logical-frame fidelity of an {\em
arbitrary} pure state $|\psi \rangle$, ${\mathbb E}\{ F_{|\psi\rangle}
(T) \}={\mathbb E}\{ |\langle \psi| \tilde{U}(T)| \psi\rangle|^2 \}$,
are given by ${\mathbb E}\{F_{|\psi\rangle}(T)\} \geq 1 - {\cal O}(T^2
T_c^2\kappa^4)$ for {\tt PDD}, and ${\mathbb
E}\{F_{|\psi\rangle}(T)\}\geq 1 - {\cal O}(T \Delta t \kappa^2)$ for
{\tt NRD}, respectively~\cite{Viola2005Random}.  Within their regime
of validity, these bounds indicate the potential for {\tt NRD} to
outperform {\tt PDD} when $|{\cal G}|^2 (T\Delta t \kappa^2) \gg 1$.
In order to quantitatively compare DD schemes, a control metric which
is both efficiently computable and state-independent is desirable.
Here, we consider {\em average} (unlike worst-case as
in~\cite{Viola2005Random}) performance, and remove the dependence upon
$|\psi\rangle$ by invoking the (expected) entanglement
fidelity~\cite{ent_fidelity}, evaluated as ${\mathbb E}\{ F_e(T)\} =
{\mathbb E}\{ |\mbox{tr}(\tilde{U}(T))/d|^2\}$. $F_e(T)$ is linearly
related to the average of $F_{|\psi\rangle} (T)$ over all
$|\psi\rangle$.  Perfect DD corresponds in this metric to ${\mathbb
E}\{F_e(T)\} \rightarrow 1$.  In practice, Monte Carlo simulations of
the controlled dynamics are used to estimate ${\mathbb E}\{F_e(T)\}$
through a statistical average $\langle \langle F_e (T)\rangle \rangle$
over control realizations.

{\em Time-independent case: Convergence improvement.}-- By way of
illustration, consider $N$ spin-1/2 particles (qubits), described by
the following model Hamiltonian:
\begin{equation}
H=\sum_{i=1}^{N} \frac{\omega_i}{2} \sigma_i^{(z)} + 
  \sum_{i<j}^N \sum_{a=x,y,z} J_{ij}^{(a)} (r_{ij})\,
 \sigma_i^{(a)} \otimes \sigma_j^{(a)} \:,
\label{ham}
\end{equation}
where $\sigma^{(x,y,z)}=X,Y,Z$ are Pauli operators, and $\omega_i,
J^{(a)}_{ij}$, denote the frequency of the $i$th qubit, and the
coupling strength of the $ij$th pair in the $a$ direction,
respectively.  Heisenberg interactions exponentially decaying with the
distance $r_{ij}$ are typical of quantum dot arrays~\cite{Loss},
whereas cubic decays describe dipolarly coupled spins in systems
ranging from nuclear magnetic resonance (NMR) crystals and
liquid-crystals~\cite{HaeberlenBook,Baugh2005} to electrons on
Helium~\cite{Dykman01}.  We assume here that $\omega_i \approx
\omega$, and work in a logical-rotating frame whereby the effective
Hamiltonian ${\tilde H}_R(t)=U^{\dagger}_c(t)U^{\dagger}_R(t)
[H(t)-\omega\sum_i Z_i/2] U_R(t)U_c(t)$, $U_R(t)=\exp [-i\omega
t\sum_{i}^{N}Z_i/2 ]$~\cite{relax,Next}.

\begin{figure}[t]
\psfrag{x}{$\hspace*{-4mm} {JT_n}$} 
\psfrag{y}{$\hspace*{-6mm}\langle\langle F_e \rangle \rangle$} 
\psfrag{z}{$\hspace*{-4mm} {Jt_n}$}
\includegraphics[width=2.7in]{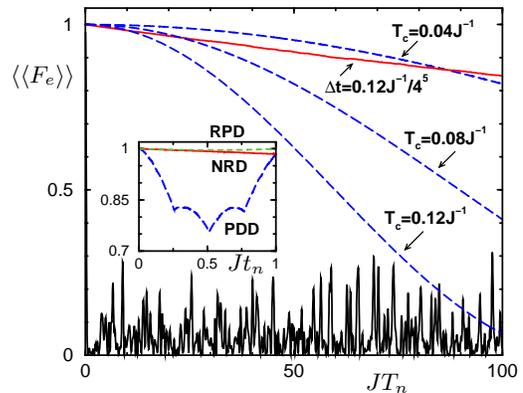}
\caption{(color online) {\tt PDD} (dashed lines) vs. {\tt NRD} (solid
line) based on a nested pulse sequence for $H$ from Eq.~(\ref{ham}) in
1D with $N=6$, $J_{ij}^{(a)}\equiv J|i-j|^{-3}$ in the
logical-rotating frame.  Main panel: average fidelity at $T_n=n|{\cal
G}| \Delta t$.  Free evolution: oscillating solid line.  Inset:
average fidelity within a cycle, $t_n=n\Delta t$, $\Delta t=10^{-3}
J^{-1}$. Dotted line: average over different path choices ({\tt RPD},
see text).  Average over $10^2$ realizations. \vspace*{-2mm}}
\label{fig:6qubits}
\end{figure}

{\tt PDD} protocols 
capable of refocusing 
$H$ for {\em arbitrary}
parameter values (and in fact, arbitrary interactions of the form
$\sigma_i^{(a)} \otimes \sigma_j^{(b)}$) may be built by recursively
nesting DD sequences based on the group ${\cal G}_i= \{{\mathbb 1}_i,
X_i, Z_i,Y_i\}$ for each added qubit, $i=2,\ldots,
N$~\cite{Stoll,Viola1,whh4}.  Although this scheme is not efficient,
as the number of pulses per cycle grows as $4^{N-1}$, it allows the
effect of large $|{\cal G}|$ to be studied in moderately small
systems.  Following ~\cite{Stoll}, the {\tt PDD} sequence we implement
corresponds to a path over the Pauli group ${\cal G}^{\tt P}
=\otimes_i {\cal G}_i$ which avoids simultaneous pulses.  In {\tt
NRD}, $U_c(t)$ is picked uniformly at random over ${\cal G}^{\tt P}$,
which may involve collective rotations on up to $R=N-1$.
qubits~\cite{SimRots}.  
Numerical results are shown in
Fig.~\ref{fig:6qubits}.  
As seen in the main panel, 
the fidelity for
{\tt NRD} decays substantially slower than for {\tt PDD}.
Irrespective of the validity of the short-time condition underlying
the lower bounds of~\cite{Viola2005Random}, these results confirm the
faster convergence of stochastic DD when $|{\cal G}|$ is large.  While
$\Delta t$ is kept fixed for {\tt NRD}, {\tt PDD} protocols with
decreasing $\Delta t$ are considered.  {\tt NRD}
eventually surpasses {all} {\tt PDD} curves, showing that, for
sufficiently long time, the constraints on $\Delta t$ for random DD
may be relaxed -- which can prove advantageous in practice.  Also
note that in situations where control constraints make it unfeasible
to complete a cycle (inset), {\tt NRD} outperforms the selected {\tt
PDD} sequence at all intra-cycle times $t_n=n\Delta t<T_c$ -- 
a finding which indicates the existence of better control paths in this
time domain.

{\em Time-independent case: Long-time improvement.}-- While different
DD protocols are equivalent in the short-time limit under ideal
averaging, performance for finite $\Delta t$ and long time depends
critically on the way in which residual errors accumulate.  We address
this point by extending our comparison to higher-level DD schemes.

In the deterministic domain, two strategies exist for improving over
{\tt PDD}. One, motivated by the Carr-Purcell sequence of NMR,
consists of a time-symmetrization of the control path, such that all
odd order terms in $\bar{H}$ are also
canceled~\cite{HaeberlenBook,Viola1}.  This leads to symmetric DD
({\tt SDD}). The other, inspired to NMR iterative
techniques~\cite{HaeberlenBook}, implements concatenated DD ({\tt
CDD})~\cite{Khodjasteh2004}.  {\tt CDD} relies on a temporal recursive
structure, the $(\ell +1)$-level pulse sequence being determined by
$C_{\ell+1}=C_{\ell} P_1 C_{\ell} P_2 \ldots C_{\ell} P_N$, where
$P_k$ is the $k$th pulse, $C_0$ is the inter-pulse interval, and $C_1$
denotes the generating {\tt PDD} sequence.

\begin{figure}[t]
\psfrag{x}{$\hspace*{-3.5mm} {JT_n}$}
\psfrag{y}{$\hspace*{-6mm}\langle \langle F_e \rangle \rangle$}
\includegraphics[width=2.55in]{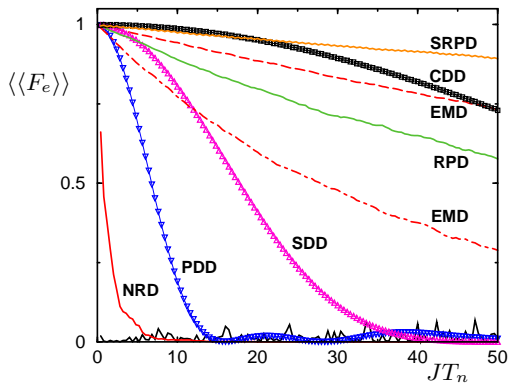}
\caption{(color online) Deterministic vs. randomized DD based on 
${\cal G}_8$
for Hamiltonian $H$ as in Fig.~\ref{fig:6qubits}, except $N=|{\cal
G}|=8$.  Data are averaged over $10^2$ control realizations, $\Delta
t= 0.05 J^{-1}$.  Note that at times $T_n=8n\Delta t$, as considered,
the inner sequences characterizing {\tt SDD}, {\tt CDD}, and {\tt
SRPD} are not necessarily completed.  For {\tt CDD}, ${\ell=5}$ is
achieved at $102.4 J^{-1}$.  Dashed and dot-dashed lines correspond to
different inner {\tt PDD} sequences for {\tt EMD}. Free evolution:
(black) solid line.
\vspace*{-3mm} }
\label{fig:ALL8}
\end{figure}

Deterministic DD is plagued by coherent error build-up at long time --
as opposed to probabilistic error accumulation in {\tt NRD}.  In order
to construct and classify stochastic DD schemes able to ensure good
performance at both short and long times, we describe each protocol in
terms of an {\em inner} and an {\em outer} code. The former determines
the control path in ${\cal G}$, thereby the pulse sequence in the
interval $[n,(n+1)]|{\cal G}|\Delta t$, subject to the condition that
a net {\em effective} Hamiltonian with norm ${\cal O}(\Delta t)$ is
obtained.  The outer code specifies additional pulses to be applied at
$T_n$ according to an outer group ${\cal G}'$, which need not coincide
with ${\cal G}$.  Combining a fixed {\tt PDD} sequence with bordering
pulses drawn at random from ${\cal G}'$ leads to {\em embedded} DD
({\tt EMD}), as in~\cite{Kern} with ${\cal G}'={\cal G}^{\tt P}$.
While choosing ${\cal G}'$ irreducible typically ensures superior
performance, {\tt EMD} schemes based on reducible ${\cal G}'={\cal G}$
may still prove useful under restricted control capabilities.
However, a potential weakness of {\tt EMD} is non-uniform performance
across the set of $|{\cal G}|!$ inner paths.  Path optimization may be
shortcut if, at every $T_n$, a random path choice is effected --
resulting in what we term {\em random path} DD ({\tt
RPD})~\cite{Viola2005Random,cyclicity}. Randomization of the inner
code ensures that DD performance are robust against path variations.
To further improve, control paths which are symmetrized in the same
manner as in {\tt SDD} may be employed, leading to 
a scheme we call
{\em symmetric random path} DD ({\tt SRPD})~\cite{kernNEW}.  
As a final step, 
{\em interpolated} DD protocols, which smoothly switch
from high-level deterministic DD (e.g. {\tt CDD}) to high-level
randomized DD ({\tt SRPD}), allow to 
optimize control performance over
the entire time axis.

Qualitative insight into the above-mentioned DD protocols may be
gained through analytical bounds on the expected fidelity decay.  For
deterministic schemes, we may write ${\mathbb E}\{F_e(T)\} \geq 1 -
{\cal O}(T^2 (|{\cal G}| \Delta t)^{2+\alpha} \kappa^{4+\alpha})$,
where $\alpha=0, 2$ for {\tt PDD}~\cite{Viola2005Random} and {\tt
SDD}~\cite{HaeberlenBook}, respectively.  Error bounds for {\tt CDD}
depend on both the system and the concatenation level, but already at
$\ell= 2$ {\tt CDD} is more efficient than {\tt SDD} in reducing
${\bar H}^{(k)}$, $k\geq2$.  Notice, however, that unlike the
single-qubit DD analyzed in~\cite{Khodjasteh2004}, no guarantee of
superpolynomial convergence exists when {\tt CDD} is based on a
reducible group.  For randomized DD, the basic error bound is
contained in~\cite{Viola2005Random} for {\tt NRD}, further adapted
in~\cite{Kern} to {\tt EMD}.  While full derivations will be presented
in~\cite{Next}, we find ${\mathbb E}\{F_e(T)\} \geq 1 - {\cal O}(T
(|{\cal G}| \Delta t)^{3+\alpha} \kappa^{4+\alpha})$, where $\alpha=0$
for {\tt EMD} and {\tt RPD}, while $\alpha=2$ for {\tt SRPD}.  As
applicability of the above bounds is strictly confined to the regime
where $T (|{\cal G}|\Delta t) \kappa^2 \ll 1$, we proceed to a
numerical comparison.

For bilinear couplings as in Eq.~(\ref{ham}), efficient combinatorial
{\tt PDD} schemes with quadratic complexity are well
known~\cite{Stoll}.  First-order DD sequences for up to $N=4^m$
qubits, $m \geq 1$, involve $N$ simultaneous pulses. For $N=8$, the
basic {\tt PDD} scheme 
is derived from the group,
\begin{eqnarray}
&&{\cal G}_8=
\{\mathbb{1},Z_3Z_4Y_5Y_6X_7X_8, \nonumber \\
&& Z_2Y_3X_4Z_6Y_7X_8,
Z_2X_3Y_4Y_5X_6Z_7, 
Y_2Y_4X_5Z_6X_7Z_8, \nonumber \\
&& Y_2Z_3X_4Z_5X_6Y_8, 
X_2Y_3Z_4X_5Z_7Y_8,X_2X_3Z_5Y_6Y_7Z_8 \}.\nonumber 
\end{eqnarray}
As ${\cal G}_8$
is reducible, saturation in the performance of {\tt CDD} is observed
when $\ell> 2$.  Numerical results are summarized in
Fig.~\ref{fig:ALL8}.  As expected, randomized DD surpasses
deterministic DD at long times, the crossing being evident between
protocols that have comparable efficiency at short times.  The
dependence of {\tt EMD} upon the underlying inner path is very
pronounced for this system.  Remarkably, {\tt CDD} is outperformed by
{\tt SRPD}.  Although {\tt CDD} is known to minimize sensitivity to
control faults not included here, preliminary results on rotation
angle imperfections indicate that the above conclusions remain
unchanged.

Exploitation of randomized DD in actual devices will vary depending on
whether probabilistic pulse generation capabilities are available or
not. In the former case, randomization appears especially promising
wherever a large number of experimental runs is required for
observation, such as in quantum dot ensemble experiments~\cite{QD}.
Even in the absence of dedicated pulsing capabilities, randomized DD
may still be useful (e.g., in solid-state NMR systems) once the
procedure is {\em de-randomized}, that is, randomization is used
off-line to post-select optimal pulse realizations for given physical
parameters.  Such an option is no longer viable for time-varying
control systems.

{\em Time-varying systems and robust performance}.-- For simplicity,
let us specialize Eq.~(\ref{ham}) to a situation where only
nearest-neighbor couplings are relevant, but a time-dependent
anisotropy is effectively present, $J_{i,i+1}^{x,y}\equiv J$,
$J^z_{i,i+1}=J\Delta (t)$. For arbitrary $N$, an efficient first-order
{\tt PDD} protocol may be constructed by alternating two collective
rotations around perpendicular axes -- one acting on all odd qubits,
the other on the even ones, e.g., if $N$ is even, ${\cal G}=\{
\mathbb{1},Z_1 Z_3 \ldots Z_{N-1}, Z_1 Y_2 Z_3 Y_4 \ldots Z_{N-1}
Y_{N}, Y_2 Y_4 \ldots Y_{N}\}$.

\begin{figure}[t]
\psfrag{x}{$\hspace*{-4mm} {JT_n}$}
\psfrag{y}{$\hspace*{-6mm}\langle \langle F_e \rangle \rangle$}
\includegraphics[width=2.8in]{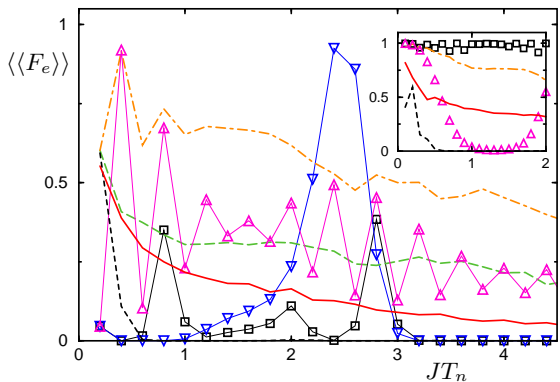}
\caption{(color online) Deterministic vs. randomized DD based on a
$|{\cal G}|=4$ sequence for 1D nearest-neighbor couplings with $N=8$
in the logical-rotating frame. $\Delta (t)=$$\sum_{k=1}^{5} \sin (10
\pi R_k J t)$, $R_k$ uniformly random in $[0.9,1.1]$.  Average
fidelity at $T_n=n|{\cal G}|\Delta t$, $10^2$ realizations.  {\tt
PDD}: 
down-triangles; {\tt SDD}: 
up-triangles, {\tt
CDD}: 
squares; {\tt NRD}: 
solid-line; {\tt RPD}: 
long-dashed line; {\tt SRPD}: 
dot-dashed line.  Free
evolution: (black) short-dashed line. Main panel: $\Delta t=0.05
J^{-1}$, $\tau/\Delta t \approx 2$.  Inset: $\Delta t= 0.025 J^{-1}$
-- notice the sensitivity of deterministic protocols (including {\tt
CDD}) to changes in $\Delta t$.\vspace*{-3mm}
}
\label{fig:NNJtime}
\end{figure}

While under mild assumptions on the time dependence of $\Delta (t)$,
the {\tt NRD} fidelity bound is still valid~\cite{Viola2005Random}, no
general result on deterministic performance is available.  Suppose
that the system fluctuation entails sign changes over time, with a
potentially unknown period $\tau$.  Then if $\tau/\Delta t \approx p$,
$p \in {\mathbb N}$, adversarial situations may arise where a
pre-established control action is inhibited or even reversed,
resulting in surprisingly poor performance of cyclic DD -- often worse
than {\tt NRD} or free evolution, see main panel of
Fig.~\ref{fig:NNJtime}.  Stochastic methods are intrinsically more
protected, {\em on average}, against such interferences, resulting in
more stable DD performance throughout.


{\em Conclusion.}-- We have shown that randomized decoupling can offer
distinctive advantages over deterministic methods in terms of faster
convergence, long-time improvement, and robust performance.  Together
with concatenation and symmetry, randomization provides a versatile
toolbox for matching different control needs.  Both the analysis of
fault-tolerance properties and of the interplay between probabilistic
design and robustness deserve a closer scrutiny, also in view of known
randomized algorithms for classical control systems~\cite{Tempo}.  It
is our hope that the present results will prompt experimental
verification in available control \vspace*{-.5mm}devices.

Partial support from Constance and Walter Burke's Special Projects
Fund in QIS and from the NSF through grant No. PHY-0555417 is
gratefully acknowledged. Thanks to O. Kern for valuable feedback.



\begin{thebibliography}{23}
\bibitem[{rel()}]{relax} \bibinfo{note}{Transforming to the rotating
frame may be avoided and/ or removing both linear and bilinear terms
is possible by slightly generalizing the schemes discussed
here~\cite{Next}.}
\bibitem[{cyc()}]{cyclicity} \bibinfo{note}{Pseudo-{\tt RPD}, where
paths always begin with ${\mathbb 1}$, is able to retain periodic overlap of
physical and logical frames, which may be important for stroboscopic
signal acquisition.  
Randomized DD requires, in general, tracking of
the applied control trajectory and further frame correction.}
\end{thebibliography}
\end{document}